\documentclass[12pt]{article}
\pdfoutput=1

\usepackage{color}
\usepackage{graphicx}
\usepackage{amsmath}
\usepackage{slashed}
\usepackage{amssymb}
\usepackage{epstopdf}

\input xy
\xyoption{all}
\xyoption{web}

\usepackage[
      colorlinks=true,
      linkcolor=blue,
      urlcolor=blue,
      filecolor=black,
      citecolor=red,
      pdfstartview=FitV,
      pdftitle={},
        pdfauthor={Michael Gutperle, Yi Li},
        pdfsubject={},
        pdfkeywords={},
        pdfpagemode={},
        bookmarksopen=true
      ]{hyperref}
\usepackage{color}
\usepackage{graphicx}

\usepackage{sectsty}
\sectionfont{\large}

\renewcommand{\thefootnote}{\fnsymbol{footnote}}
\renewcommand{\thanks}[1]{\footnote{#1}}
\newcommand{\starttext}{
\setcounter{footnote}{0}
\renewcommand{\thefootnote}{\arabic{footnote}}}

\newcommand{\bea}{\begin{eqnarray}}
\newcommand{\eea}{\end{eqnarray}}
\newcommand{\be}{\begin{equation}}
\newcommand{\ee}{\end{equation}}

 \newcommand{\reals}{\mathbb{R}}

\usepackage{ dsfont } %for identity matrix

\long\def\symbolfootnote[#1]#2{\begingroup%
\def\thefootnote{\fnsymbol{footnote}}\footnote[#1]{#2}\endgroup}

\begin{document}
\setlength{\baselineskip}{16pt}

\starttext
\setcounter{footnote}{0}

%\begin{flushright}
%August 2nd, 2014
%\end{flushright}
%
\bigskip

\begin{center}

{\Large \bf  Higher Spin Chern-Simons Theory and the Super Boussinesq hierarchy}

\vskip 0.4in

{\large  Michael Gutperle and Yi Li}

\vskip .2in

{\it Mani L. Bhaumik Institute for Theoretical Physics}\\
{ \it Department of Physics and Astronomy }\\
{\it University of California, Los Angeles, CA 90095, USA}\\[0.5cm]
\href{mailto:yli@physics.ucla.edu}{\texttt{yli@physics.ucla.edu}}\texttt{, }\href{mailto:gutperle@physics.ucla.edu}{\texttt{gutperle@physics.ucla.edu}}

\bigskip

\bigskip

\end{center}

\begin{abstract}

%\vskip 0.1in
\setlength{\baselineskip}{18pt}

In this paper we construct a map between a solution of supersymmetric Chern-Simons higher spin gravity based on the superalgebra $sl(3|2)$ with Lifshitz scaling and the $N=2$ super Boussinesq hierarchy.  We show that under this map  the time evolution equations of both theories coincide.  In addition, we identify the Poisson structure of the Chern-Simons theory induced by gauge transformation with the second Hamiltonian structure of the super Boussinesq hierarchy.
\end{abstract}

\setcounter{equation}{0}
\setcounter{footnote}{0}

%\newpage
%
%
%
%\newpage
%\tableofcontents
%
%
\newpage

\baselineskip 16pt

%%%%%%%%%%%%%%%%%%%%%%%%%%%%%%%%%%%%%%%%%%%%
%%%%%%%%%%%%%%%%%%%%%%%%%%%%%%%%%%%%%%%%%%%%
\section{Introduction}
\setcounter{equation}{0}
\label{sec1}
%%%%%%%%%%%%%%%%%%%%%%%%%%%%%%%%%%%%%%%%%%%%
%%%%%%%%%%%%%%%%%%%%%%%%%%%%%%%%%%%%%%%%%%%%

The Chern-Simons formulation of three dimensional higher spin gravity has been a subject of great activity  recent years. The interest was started by the discovery that Chern- theories  in three space-time dimensions based on gauge algebras such as  $sl(N,\mathbb{R})$ and  $hs(\lambda)$ \cite{Blencowe:1988gj,Bergshoeff:1989ns} are versions of Vasiliev higher spin theories \cite{Vasiliev:2000rn,Vasiliev:1999ba}.  Furthermore the Chern-Simons theories can realize the asymptotic symmetries of $W_N$ CFT's   \cite{Henneaux:2010xg,Campoleoni:2010zq}
 which have conserved currents of spin greater than two.  Gaberdiel and Gopakumar    \cite{Gaberdiel:2010pz,Gaberdiel:2012uj} proposed a concrete holographic duality of higher spin Chern-Simons theory and $W_N$ minimal models which provides a  new arena to test ideas of higher spin gravity and AdS/CFT.

The Chern-Simons higher spin gravity has also been used to construct solutions which are not asymptotically Anti-de Sitter but obey other asymptotics, such as Schroedinger, warped AdS or Lifshitz geometries (see e.g.  \cite{Gary:2012ms,Afshar:2012nk,Gonzalez:2013oaa,Gutperle:2013oxa}). In two papers  \cite{Gutperle:2014aja,Beccaria:2015iwa} the authors have investigated asymptotically Lifshitz solution of Chern-Simons  higher spin gravity and found an intriguing relation to integrable systems. In particular, it was shown that there exists an explicit map between the asymptotic Lifshitz solution of $sl(N,\mathbb{R})$ Chern-Simons theory and (integer) Lifshitz scaling exponent $z$  to the $(m,n)$ member of the KdV hierarchy, where the parameters of both theories are identified as $m=z$ and $n=N$.

The Chern-Simons higher spin theories based on the Lie algebras $sl(N,\mathbb{R})$ and $hs(\lambda)$ are purely bosonic theories, with higher spin fields of integer spin.  A supersymmetric generalization of the bosonic theories can be achieved by considering Chern-Simons theories based on Lie superalgebras such as $sl(n,m)$, see e.g  \cite{Candu:2012jq,Henneaux:2012ny,Tan:2012xi}. The goal of the present paper is to investigate the relation between the Chern-Simons higher spin theories and integrable systems for supersymmetric theories. In particular we focus on one of the simplest examples, Chern-Simons theory based on the Lie superalgebra $sl(3|2)$.

The structure of the paper is as follows: In section \ref{sec2} we review the KdV hierarchy and its relation to higher spin Chern-Simons theory given in our previous work  \cite{Gutperle:2014aja,Beccaria:2015iwa}. In section \ref{sec3} we review the supersymmetric generalization of a particular member of the KdV hierarchy, namely the $N=2$ super Boussinesq hierarchy. In section \ref{sec4} we construct a supersymmetric higher spin Chern-Simons theory based on the Lie superalgebra $sl(3|2)$ which enjoys Lifshitz symmetry and provide an explicit map between this theory and the $N=2$ super Boussinesq  hierarchy such that the time evolution equations coincide. In section \ref{sec5} we explain where this correspondence  comes from by showing that the Poisson structure of both theories can be mapped into each other. We close with a discussion of possible directions for future research in section \ref{sec6}.

%%%%%%%%%%%%%%%%%%%%%%%%%%%%%%%%%%%%%%%%%%%%
%%%%%%%%%%%%%%%%%%%%%%%%%%%%%%%%%%%%%%%%%%%%
\section{KdV hierarchy and the Higher-Spin Chern Simons theory}
\setcounter{equation}{0}
\label{sec2}
%%%%%%%%%%%%%%%%%%%%%%%%%%%%%%%%%%%%%%%%%%%%
%%%%%%%%%%%%%%%%%%%%%%%%%%%%%%%%%%%%%%%%%%%%

The $n$-th KdV hierarchy is a bi-Hamiltonian integrable system of $n-1$ fields $u_2,u_3,\cdots, u_n$ with commuting Hamiltonian flows, where the second Hamiltonian structure is  the $W_n$ algebra \cite{gelfand:1976fp,dickey:1997lc}. It's conveniently formulated by  utilizing the formalism of pseudo-differential operators
\be\label{ldefa}
 L=\partial^n + u_2 \partial^{n-2} + \ldots + u_{n-1} \partial + u_n
\ee
and the $m$-th Hamiltonian flow is given by the Lax-type equation \cite{dickey:1997lc,battle:ln}
\be
 \dot{L}=[L^\frac{m}{n}_+,L]
\ee
For any of the Hamiltonian flows, an infinite tower of conserved quantities exist
\be
 Q_k=\int dx\;  {\rm res} L^\frac{k}{n},\quad k \in \mathbb{N}
\ee
where ${\rm res}$ denotes the residue of the pseudo-differential operator, that is, the coefficient of $\partial^{-1}$.
This equation is invariant under Lifshitz scaling of the space and time coordinates with scaling exponent $m$
\be
 x \rightarrow \lambda x, \quad t \rightarrow \lambda^m t
\ee
where the scaling of the fields $u_i$ is determined by dimensional analysis from (\ref{ldefa}).

The Chern-Simons action at level $k$ in three dimensional space-time is  given by the following action
\be\label{chernsimonsa}
S_{CS}[A]= {k\over 4\pi}  \int {\rm tr}\Big( A\wedge dA+{2\over 3} A\wedge A\wedge
A\Big)
\ee
where $A$ is a Lie algebra valued gauge connection. It was shown in \cite{Blencowe:1988gj,Bergshoeff:1989ns} that higher spin gravity theory in three dimensions can be formulated by combining two copies of Chern-Simons actions with level $k$ and $-k$ respectively, and with gauge algebra $sl(N,\reals)$ or $hs(\lambda)$.
\be
S= S_{CS}[A]-S_{CS}[\bar A]
\ee
The action yields the equation of motion, also known as the flatness condition
\be\label{flatness}
F=dA+ A\wedge A=0, \quad  \bar F=d\bar A+\bar A\wedge \bar A=0
\ee
The relation to higher spin gravity is made by expressing Lie algebra valued generalizations of the vielbein and spin connection in terms of the gauge connections
\be
e_\mu ={1\over 2} (A_\mu -\bar A_\mu), \quad \omega_\mu ={1\over 2} (A_\mu + \bar A_\mu)
\ee
where a unit length has been chosen for dimensional match.
The metric is then given by
\be
g_{\mu\nu}={1\over {\rm tr}(L_0)^2} {\rm tr}(e_\mu e_\nu)
\ee
where $L_0$ is the Cartan generator of $sl(2,\mathbb{R})$ subalgebra of $sl(N,\mathbb{R})$ or $hs(\lambda)$. The higher spin fields of Vasiliev theory can be expressed in terms of  higher order traces involving  the generalized vielbein $e_\mu$.

In the context of holography a "radial coordinate" $\rho$ is introduced where the holographic boundary is defined at $\rho\to \infty$. In addition we define a time-like coordinate $t$ and a space-like coordinate $x$. The asymptotic Lifshitz metric with Lifshitz scaling exponent $z$ is defined as the metric that takes the form
\be
ds^2\sim d\rho^2- e^{2z \rho} dt^2+ e^{2\rho} dx^2
\ee
as $\rho \rightarrow \infty$.
By the choice of the radial gauge \cite{Gutperle:2014aja} we single out the $\rho$ dependence of the connection
\bea\label{radial}
 A_\mu(\rho,x,t)&=&b(\rho)^{-1} a_\mu(x,t) b(\rho) + b(\rho)^{-1} \partial_\mu b(\rho) \nonumber \\ \bar{A}_\mu(\rho,x,t)&=&b(\rho) \bar{a}_\mu(x,t) b(\rho)^{-1} + b(\rho) \partial_\mu b(\rho)^{-1}
\eea
where $b(\rho)=e^{\rho L_0}$ and $a_\rho=\bar{a}_\rho=0$, then the flatness conditions (\ref{flatness}) are  reduced to
equations of $\rho$ independent connections $a_x,a_t$ and $\bar{a}_x,\bar{a}_t$
\be\label{flat}
\partial_t a_x -\partial_x a_t + [a_t,a_x]=0, \quad \partial_t \bar{a}_x -\partial_x \bar{a}_t + [\bar{a}_t,\bar{a}_x]=0
\ee
An infinitesimal gauge transformation generated by a parameter $\Lambda$ is
\be\label{gaugetrans}
\delta_\Lambda A = d\Lambda + [A,\Lambda]
\ee
The gauge transformation preserves the radial gauge when the parameter takes the form $\Lambda(\rho,x,t)=b(\rho)^{-1} \lambda(x,t) b(\rho)$, and it translates to gauge transformation on the $\rho$-independent connection as
\be\label{gaugetrans1}
 \delta_\lambda a = da + [a,\lambda]
\ee
Moreover, we take $a_t$ as an differential polynomial in $a_x$, thus the flatness condition becomes a time evolution equation of  $a_x$, and the time evolution is essentially a gauge transformation with a field dependent gauge transformation parameter $\lambda =a_t$.

For the gauge algebra $sl(N,\reals)$, we can put $a_x$ in the lowest weight gauge, in which only the $N-1$ lowest weight terms in $sl(N,\reals)$ are dynamical. We then construct $a_t$ as a differential polynomial of $a_x$ that generates  an asymptotic Lifshitz spacetime with $a_x$ and keeps $a_x$ in the lowest weight gauge under time evolution. These requirements do not fix $a_t$ completely, and we get a time evolution equation of $N-1$ fields in $a_x$ with undetermined coefficients depending on the choice of $a_t$. The solution leads to an asymptotically Lifshitz spacetime and hence duals a two dimensional field theory living on the boundary that has Lifshitz scaling symmetry.

The relation between the KdV hierarchy and the Lifshitz Chern-Simons theory was first motivated by the observation in \cite{Gutperle:2013oxa} that the  equation of motion of $sl(3,\reals)$ $z=2$ Lifshitz Chern-Simons theory takes the form of the Boussinesq equation.
 Further investigation  reveals a closer relation between the two theories, most importantly they both possess Lifshitz scaling symmetry and can both be brought in a  Lax type equation form. In our previous work \cite{Gutperle:2014aja,Beccaria:2015iwa}, it was shown that by suitable choice of $a_t$ called the "KdV gauge", the asymptotic Lifshitz solution with Lifshitz scaling coefficient $z$ of $sl(N,\reals)$ Chern-Simons theory can be identified with the $(n,m)$ member of the KdV hierarchy where $n=N$ and $m=z$ by an explicit map.

%%%%%%%%%%%%%%%%%%%%%%%%%%%%%%%%%%%%%%%%%%%%
%%%%%%%%%%%%%%%%%%%%%%%%%%%%%%%%%%%%%%%%%%%%
\section{N=2 Super Boussinesq hierarchy}
\setcounter{equation}{0}
\label{sec3}
%%%%%%%%%%%%%%%%%%%%%%%%%%%%%%%%%%%%%%%%%%%%
%%%%%%%%%%%%%%%%%%%%%%%%%%%%%%%%%%%%%%%%%%%%

The main goal of this paper is to extend the established relation between the Lifshitz Chern-Simons theory and the KdV hierarchy to the supersymmetric case. For the Lifshitz Chern-Simons theory, a supersymmetric extension can be naturally constructed  by replacing the Lie algebra by a Lie superalgebra which contains the Lie algebra as a bosonic subalgebra. It is clear however that this extension is not unique. On the other hand, the supersymmetric extension of the KdV hierarchy is not as well studied  as the bosonic case and no general classification  exists to our knowledge. 

Therefore, instead of pursuing  a general construction,  we restrict ourselves to a concrete and workable example in the present paper. In particular, we want to find the supersymmetric extension of the correspondence between $sl(3,\reals)$ $z=2$ Lifshitz Chern-Simons theory and $(n=3,m=2)$ member of  the KdV hierarchy. The Lie superalgebra $sl(3|2)$ is a natural extension of $sl(3,\reals)$, and $sl(3|2)$ Chern-Simons theory has been  studied in the past, see for example  \cite{Peng:2012ae,Chen:2013oxa}. Because $sl(3|2)$ has two sets of fermionic generators, we should look for $N=2$ supersymmetric extension of $n=3$ KdV hierarchy, that is, $N=2$ super Boussinesq hierarchy. Since the second Hamiltonian structure of the Boussinesq hierarchy is the $W_3$ algebra, one should expect $N=2$ super Boussinesq hierarchy to possess $N=2$ super $W_3$ algebra as the second Hamiltonian structure. Guided by this principle, $N=2$ super Boussinesq hierarchy was constructed in \cite{Bellucci:1993wq,Ivanov:1992tp} in terms of two bosonic superfields $J$ and $T$ in the superspace  coordinates $(x,\theta,\bar{\theta})$
\begin{align}
&J(x,\theta,\bar{\theta})=\bar{\theta}\theta u(x) + \theta \xi(x) + \bar{\theta} \bar{\xi}(x) + y(x) \nonumber \\
&T(x,\theta,\bar{\theta})=\bar{\theta}\theta z(x) + \theta \eta(x) + \bar{\theta} \bar{\eta}(x) + v(x)
\end{align}
with two free parameters $c$ and $\alpha$, where $c$ is a free constant in the $N=2$ super $W_3$ algebra realized by $J,T$ that corresponds to rescaling freedom of $J,T$, and $\alpha$ is a free constant in the Hamiltonian $H=\int dx d\theta d\bar{\theta} \; (T+\alpha J^2)$ that generates the time evolution
\be
 \dot{J}=\{J,H\}, \quad \dot{T}=\{T,H\}
\ee
The super Boussinesq equation should reduce to the Boussinesq equation when $sl(3|2)$ reduces to $sl(3,R)$, and that's possible only when the parameter $\alpha$ takes the following value $\alpha=-\frac{4}{c}$ \cite{Ivanov:1992tp}. After setting $c=-\frac{4}{\alpha}$, the $N=2$ super Boussinesq equation reads in terms of superfields
\bea\label{jtevolve}
 \dot{J}&=&2T^{'}-\delta J^{'} + 4\alpha J J^{'} \nonumber 
\\
 \dot{T}&=&-2J^{'''} + \delta T^{'} - 20\alpha \partial(\bar{D}JDJ) + 8\alpha J^{'}\delta J + 4\alpha J \delta  J^{'} \nonumber \\
 &+& 16 \alpha^2 J^2 J^{'} -12 \alpha \bar{D}J DT -12 \alpha DJ \bar{D}T -12 \alpha J^{'}T -4\alpha J T^{'}
\eea
where
\bea
 D&=&\partial_{\theta} - \frac{1}{2}\bar{\theta}\partial, \quad \bar{D}=\partial_{\bar{\theta}} - \frac{1}{2}\theta\partial \\
 \delta &=& [\bar{D},D]
\eea

It was shown in \cite{Bellucci:1993wq} that if we choose $c=8$ the parameter $\alpha$ must take one of these three values $-2,-\frac{1}{2},\frac{5}{2}$ for the equation to be integrable in the sense that higher order conserved charges exist. We see $\alpha=-\frac{4}{c}=-\frac{1}{2}$ is indeed one of them, and later an elegant Lax pair formulation of this case was given in \cite{Delduc:1996mx}. In the form in components the time evolution equations (\ref{jtevolve}) read
\begin{align}\label{boussa}
 &\dot{y}=2(u+v)^{'}+4\alpha yy^{'} \nonumber\\
 &\dot{\xi}=\xi^{''}+2\eta^{'}+4\alpha(y\xi)^{'} \nonumber \\
 &\dot{\bar{\xi}}=-\bar{\xi}^{''}+2\bar{\eta}^{'}+4\alpha(y\bar{\xi})^{'} \nonumber\\
 &\dot{u}=2z^{'}+\frac{1}{2} y^{'''} + 4\alpha (yu)^{'} + 4\alpha (\xi\bar{\xi})^{'} \nonumber\\
 &\dot{v}=-2z^{'}-16\alpha u y^{'} -8\alpha u^{'}y - 4\alpha y v^{'} - 12\alpha y^{'} v + 12 \alpha (\eta\bar{\xi}+\bar{\eta}\xi) + 20\alpha (\xi\bar{\xi})^{'} - 2y^{'''} + 16\alpha^2 y^2y^{'}\nonumber\\
 &\dot{\eta}= - \eta^{''} - 2\xi^{'''} - 28\alpha u^{'} \xi - 36\alpha u \xi^{'} - 10\alpha v^{'} \xi - 12\alpha v \xi^{'} - 12 \alpha u\eta + 12\alpha z\xi \nonumber\\
 &\;\;\;+ 10\alpha y^{''} \xi + 32\alpha^2 yy^{'}\xi + 2\alpha y^{'}\xi^{'} + 16\alpha^2 y^2\xi^{'} - 4\alpha y\xi^{''} - 6\alpha y^{'}\eta - 4\alpha y\eta^{'} \nonumber\\
 &\dot{\bar{\eta}}= \bar{\eta}^{''} - 2\bar{\xi}^{'''} -28\alpha u^{'} \bar{\xi} -36\alpha u \bar{\xi}^{'} - 10\alpha v^{'}\bar{\xi} - 12\alpha v\bar{\xi}^{'} + 12\alpha u\bar{\eta} - 12\alpha z\bar{\xi} \nonumber\\
 &\;\;\;-10\alpha y^{''} \bar{\xi} + 32\alpha^2 yy^{'}\bar{\xi} - 2\alpha y^{'}\bar{\xi}^{'} + 16\alpha^2 y^2 \bar{\xi}^{'} + 4\alpha y\bar{\xi}^{''} - 6\alpha y^{'}\bar{\eta} -4\alpha y\bar{\eta}^{'} \nonumber\\
 &\dot{z}= - 2u^{'''} - \frac{1}{2}v^{'''} - 64\alpha uu^{'} -16\alpha uv^{'} - 12\alpha u^{'}v + 32\alpha^2 yy^{'}u + 16\alpha^2 y^2u^{'} - 4\alpha yz^{'} - 2\alpha yy^{'''} \nonumber \\
 &\;\;\;+ 6\alpha y^{'}y^{''} + 10\alpha \bar{\xi} \eta^{'} + 6\alpha \bar{\xi}^{'} \eta - 10\alpha \xi\bar{\eta}^{'} - 6\alpha \xi^{'}\bar{\eta} + 14\alpha \xi\bar{\xi}^{''} - 14\alpha \xi^{''}\bar{\xi} + 32\alpha^2 (y\xi\bar{\xi})^{'} 
\end{align}

%%%%%%%%%%%%%%%%%%%%%%%%%%%%%%%%%%%%%%%%%%%%
%%%%%%%%%%%%%%%%%%%%%%%%%%%%%%%%%%%%%%%%%%%%
\section{$sl(3|2)$ Lifshitz Chern-Simons theory and the map to $N=2$ super Boussinesq hierarchy}
\setcounter{equation}{0}
\label{sec4}
%%%%%%%%%%%%%%%%%%%%%%%%%%%%%%%%%%%%%%%%%%%%
%%%%%%%%%%%%%%%%%%%%%%%%%%%%%%%%%%%%%%%%%%%%
When we take the gauge algebra to be Lie superalgebra, most notably $sl(p|q)$, the Chern-Simons action takes the form
\be\label{schernsimonsa}
S_{CS}[A]= {k\over 4\pi}  \int {\rm str}\Big( A\wedge dA+{2\over 3} A\wedge A\wedge
A\Big)
\ee
where $\rm str$ denotes the supertrace. In complete analogy to the non-supersymmetric case, higher spin supergravity  can be formulated by two copies of Chern-Simons actions, with the vielbein and spin connection expressed in terms of the gauge connection
\be
e_\mu ={1\over 2} (A_\mu -\bar A_\mu), \quad \omega_\mu ={1\over 2} (A_\mu + \bar A_\mu)
\ee
and the metric is given by
\be\label{metric}
g_{\mu\nu}={1\over {\rm str}(L_0)^2} {\rm str}(e_\mu e_\nu) = {1\over 4{\rm str}(L_0)^2} {\rm str}((A_\mu -\bar A_\mu)(A_\nu -\bar A_\nu))
\ee
Now we focus on $sl(3|2)$ Lifshitz Chern-Simons theory, that is, Chern-Simons theory with $sl(3|2)$ gauge algebra that gives asymptotic Lifshitz spacetime. Background material on  $sl(3|2)$ relevant to this paper are reviewed in Appendix A. We follow the notation of generators of $sl(3|2)$ in \cite{Chen:2013oxa} and the super matrix representation which we include for completeness can also be found there. We adopt the radial gauge as we did in the non-supersymmetric case
\be
 A_\mu(\rho,x,t)=b(\rho)^{-1} a_\mu(x,t) b(\rho) + b(\rho)^{-1} \partial_\mu b(\rho), \quad \bar{A}_\mu(\rho,x,t)=b(\rho) \bar{a}_\mu(x,t) b(\rho)^{-1} + b(\rho) \partial_\mu b(\rho)^{-1}
\ee
where $b(\rho)=e^{\rho L_0}$ and $a_\rho=\bar{a}_\rho=0$. Clearly the weight of terms in $a_\mu$ will translate to growth rate with $\rho$ in $A_\mu$ because the weight is the eigenvalue of the commutator with $L_0$. An exact Lifshitz spacetime can be obtained by setting
\bea
 &a_x=L_1, \quad a_t= \frac{\sqrt{3}}{4} W_2 \\
 &\bar{a}_x= L_{-1}, \quad \bar{a}_t= \frac{\sqrt{3}}{4} W_{-2}
\eea
that is
\bea
 &A= L_0 d\rho + L_1 e^\rho dx + \frac{\sqrt{3}}{4} W_2 e^{2\rho} dt \\
 &\bar{A}= -L_0 d\rho + L_{-1} e^\rho dx + \frac{\sqrt{3}}{4} W_{-2} e^{2\rho} dt
\eea
One can verify that by (\ref{metric}) the connection yields Lifshitz spacetime $ds^2=d\rho^2+e^{2\rho}dx^2-e^{4\rho}dt^2$ with Lifshitz scaling exponent $z=2$. Now we add dynamical terms to the connection but keeping the leading term fixed to get asymptotic Lifshitz spacetime. We will focus on the unbarred sector here, the barred sector can be worked out by the same algorithm thanks to the weight flipping automorphism of $sl(3|2)$. The ansatz of $a_x$ in the lowest weight gauge is
\begin{align}
 a_x = L_1 + j J + a A_{-1} + l L{-1} + w W_{-2} + g G_{-\frac{1}{2}} + h H_{-\frac{1}{2}} + s S_{-\frac{3}{2}} +t T_{-\frac{3}{2}}
\end{align}
with all the dynamical terms being the lowest weight elements in $sl(3|2)$. The component  $a_t$ should start with $\frac{\sqrt{3}}{4}W_2$, and its non-highest weight terms are completely determined by highest weight terms because it must preserve the lowest weight gauge of $a_x$ in time evolution. We take the highest weight terms to be differential polynomials of fields in $a_x$ of the correct dimension, so $a_t$ must take the form
\begin{align}
 a_t = \frac{\sqrt{3}}{2}(\frac{1}{2} W_2 + (d_1 a + d_2 l + d_3 j^2 + d_4 j^{'}) J + c_1 j A_1 + c_2 j L_1 + c_3 g G_\frac{1}{2} + c_4 h H_\frac{1}{2} + \ldots)
\end{align}
with eight free constants $c_1,c_2,c_3,c_4,d_1,d_2,d_3,d_4$, and with non-highest weight terms omitted. We factor out $\frac{\sqrt{3}}{2}$ for calculational simplicity. Now we deal with the problem of fixing $a_t$ to map the time evolution equation of $a_x$ to the $N=2$ super Boussinesq equation. In order to have the lowest dimensional conserved bosonic charge and fermionic quantities, $\dot{j}$ and $\dot{g}$ must be total derivatives. This condition fixes $a_t$ up to only one free constant $c_3$
\begin{align}
 &c_1=-1 \nonumber \\
 &c_2=-2c_3 +\frac{5}{3}  \nonumber \\
 &c_4=-c_3 \nonumber \\
 &d_1=\frac{1}{9}(8-15c_3)\nonumber \\
 &d_2=-c_3 \nonumber \\
 &d_3=-3c_3 \nonumber \\
 &d_4=0
\end{align}
It turns out if we set $c_3=\frac{1}{3}$, the time evolution equation of $a_x$ can be identified with the $N=2$ super Boussinesq equation after we rescale the time evolution by a factor $-2\sqrt{3}$, that is equivalent to replacing $a_t$ by $-2\sqrt{3}a_t$. The time evolution equation of $a_x$ we get after replacing $a_t$ by $-2\sqrt{3}a_t$ is
\begin{align}
 &\dot{j}= (l-a+3j^2)^{'}\nonumber  \\
 &\dot{g}= g^{''} - 6s^{'} + 4(jg)^{'} \nonumber  \\
 &\dot{h}= -h^{''} - 6t^{'} + 4(jh)^{'}\nonumber  \\
 &\dot{a}= \frac{3}{2}j^{'''}+6w^{'}+3jl^{'}+6lj^{'}-6aj^{'}-3ja^{'}+\frac{15}{2}(gh)^{'}+\frac{27}{2}(gt+hs)\nonumber  \\
 &\dot{l}= -\frac{3}{2}j^{'''}+6w^{'}-3jl^{'}-6lj^{'}+6aj^{'}+3ja^{'}-\frac{33}{2}(gh)^{'}-\frac{45}{2}(gt+hs)\nonumber  \\
 &\dot{s}= -s^{''} + \frac{2}{3} g^{'''} - (10a+6l+6j^2+6j^{'})s -4js^{'} +(\frac{10}{3}a^{'}+\frac{14}{3}l^{'}-\frac{10}{3}j^{''}-16w+\frac{32}{3}aj-\frac{4}{3}jj^{'})g\nonumber \\
 &\;\;+ (\frac{14}{3}a+\frac{2}{3}j^2+6l-\frac{2}{3}j^{'})g^{'} + \frac{4}{3}jg^{''} \nonumber  \\
 &\dot{t}= t^{''}+\frac{2}{3} h^{'''}+(10a+6l+6j^2-6j^{'})t-4jt^{'}+(\frac{10}{3}a^{'}+\frac{14}{3}l^{'}+\frac{10}{3}j^{''}+16w-\frac{32}{3}aj-\frac{4}{3}jj^{'})h \nonumber \\
 &\;\;+ (\frac{14}{3}a+\frac{2}{3}j^2+6l+\frac{2}{3}j^{'})h^{'} - \frac{4}{3}jh^{''}\nonumber \\
 &\dot{w}=-\frac{1}{4}(a+l)^{'''}-2[(a+l)^2]^{'}-4j^{'}gh+j(gh)^{'}+9j(gt+hs)-\frac{7}{4}(gh^{''}+hg^{''})\nonumber  \\
 &\;\;-\frac{9}{4}g^{'}t-\frac{15}{4}gt^{'}+\frac{9}{4}h^{'}s+\frac{15}{4}hs^{'}
\end{align}
which can be identified with the $N=2$ super Boussinesq equation via the explicit map
\begin{align}
 &j=\alpha y \nonumber  \\
 &g=k\xi, \quad h=-k\bar{\xi} \nonumber \\
 &a=-\frac{3}{4}\alpha v, \quad l=-\alpha^2y^2+\frac{5}{4}\alpha v +2\alpha u \nonumber  \\
 &s=-\frac{1}{3} k \eta, \quad t=\frac{1}{3} k \bar{\eta} \nonumber \\
 &w=\frac{\alpha}{4}z-\frac{\alpha^2}{2}yv
\end{align}
with $k^2=2\alpha^2$, no matter which root $k$ takes.

%%%%%%%%%%%%%%%%%%%%%%%%%%%%%%%%%%%%%%%%%%%%
%%%%%%%%%%%%%%%%%%%%%%%%%%%%%%%%%%%%%%%%%%%%
\section{The Poisson structure of Lifshitz Chern-Simons theory and the second Hamiltonian structure of  the $N=2$ super Boussinesq hierarchy}
\setcounter{equation}{0}
\label{sec5}
%%%%%%%%%%%%%%%%%%%%%%%%%%%%%%%%%%%%%%%%%%%%
%%%%%%%%%%%%%%%%%%%%%%%%%%%%%%%%%%%%%%%%%%%%
In the previous sections we have worked out a specific example of the relation between supersymmetric  Chern-Simons   Lifshitz theory and  super Boussinesq  hierarchy, that is, we established the map between $sl(3|2)$ Lifshitz Chern-Simons theory and $N=2$ super Boussinesq hierarchy such that the time evolution equations of the two theories coincide. In this section we argue  that there is a structurally deeper connection of the two theories. In the following we will show that the Poisson structure of $sl(3|2)$ Lifshitz Chern-Simons theory induced by gauge transformation is identical to the second Hamiltonian structure of $N=2$ super Boussinesq hierarchy.

The time evolution of Chern-Simons is essentially a gauge transformation with gauge transformation parameter $a_t$, that is $\dot{a}_x=\delta_{a_t}a_x=\partial_x a_t + [a_x,a_t]$. Fixed in the lowest weight gauge, the gauge transformation induces a Poisson structure of the fields in the reduced phase space \cite{Campoleoni:2010zq}. That is, the gauge transformation of a field $\phi$ with gauge parameter $\lambda$ is regarded as a Poisson bracket between the field and and the charge associated with the gauge transformation parameter
\begin{align}
 &\delta_\lambda \phi = \{Q_\lambda,\phi\}
\end{align}
where the charge is given by
\begin{align}
 \delta Q_{\lambda}=C \int dx \; {\rm str} \;\lambda \delta a_x
\end{align}
with $C$ being an arbitrary constant. The Poisson brackets of all fields can be computed by choosing different gauge parameters, and it's used to calculate the boundary charge algebra in the context of holography, see for example  \cite{Peng:2012ae}. Now the time evolution equation of Chern-Simons theory can be recast in a form resembling the Hamiltonian dynamics
\begin{align}
 \dot{a_x}=\{Q_{a_t},a_x\}
\end{align}
On the other hand, the time evolution of the $N=2$ super Boussinesq hierarchy is  generated by its Hamiltonian structure
\begin{align}
 \dot{T}=\{T,H\}, \quad \dot{J}=\{J,H\} 
\end{align}
Since we have a map between the two theories that identifies the time evolution equation, it's natural to conjecture the Poisson structure of the $sl(3|2)$ Chern-Simons theory is identical to the second Hamiltonian structure of $N=2$ super Boussinesq hierarchy via the established map. Note we have replaced $a_t$ by $-2\sqrt{3}a_t$ to make the map, it's actually
\begin{align}
 \dot{a_x}=\{-2\sqrt{3}Q_{a_t},a_x\}
\end{align}
that is identified with the $N=2$ super Boussinesq equation, therefore we must have $2\sqrt{3}Q_{a_t}=H$. Straightforward computation yields
\begin{align}
 2\sqrt{3}Q_{a_t}=2C \int dx j(l-a)+4w+j^3-gh=6\alpha C \int dx z+2\alpha(yu+\xi\bar{\xi})
\end{align}
where we have used the map between two theories. On the other hand, the Hamiltonian of the second Hamiltonian structure of $N=2$ super Boussinesq hierarchy is given as
\begin{align}
 H=\int dx d\theta d\bar{\theta} (T+\alpha J^2)=\int dx z+2\alpha(uy+\xi\bar{\xi})
\end{align}
We see that $2\sqrt{3}Q_{a_t}$ is equal to the Hamiltonian in the second Hamiltonian structure of the $N=2$ super Boussinesq hierarchy with the choice $C=\frac{1}{6\alpha}$. We have computed the Poisson structure of the $sl(3|2)$ Chern-Simons theory with $C=\frac{1}{6\alpha}$ and listed in Appendix B. One can verify it's indeed identical to the second Hamiltonian structure of the $N=2$ super Boussinesq hierarchy given in \cite{Bellucci:1993wq}.

%%%%%%%%%%%%%%%%%%%%%%%%%%%%%%%%%%%%%%%%%%%%
%%%%%%%%%%%%%%%%%%%%%%%%%%%%%%%%%%%%%%%%%%%%
\section{Discussion}
\setcounter{equation}{0}
\label{sec6}
%%%%%%%%%%%%%%%%%%%%%%%%%%%%%%%%%%%%%%%%%%%%
%%%%%%%%%%%%%%%%%%%%%%%%%%%%%%%%%%%%%%%%%%%%
In this paper we worked out a concrete example for an extension of the relation between the Lifshitz Chern-Simons theory and the KdV hierarchy to the supersymmetric case. It was shown  that $sl(3|2)$ Lifshitz Chern-Simons theory, as the supersymmetric extension of $sl(3,\reals)$ Lifshitz Chern-Simons theory, corresponds to $N=2$ super Boussinesq hierarchy constructed in \cite{Bellucci:1993wq} with the appropriate choice of parameters $\alpha=-\frac{4}{c}$.   It was found in \cite{Bellucci:1993wq} that for $c=8$ there are three values of $\alpha$ (including the one we choose $\alpha=-\frac{1}{2}$) such that the equation obtained is an integrable system,  in the sense that  an infinite tower  of higher order conserved quantities exist. In addition for some of them a Lax pair formulation exists or a bi-Hamiltonian structure exists \cite{Bellucci:1993wq,Delduc:1996mx}.

The results in the present paper leave some questions open of  which we list a few here:

 It is a natural question to ask if the super Boussinesq hierarchy with other values of the parameter $\alpha$ also corresponds to Lifshitz Chern-Simons theory with other gauge algebra different from  $sl(3|2)$.   In fact, in almost all the cases of supersymmetric extension of KdV hierarchies, it turns out we have to choose a discrete set of values of the parameters to make the theory integrable \cite{Labelle:1990vv,Delduc:1993sk,Delduc:1995ad}. If we can formulate all these supersymmetric extensions of KdV by Lifshitz Chern-Simons theory with different Lie superalgebras, we may be able to explain the choices of discrete values of parameters in the perspective of the theory of Lie superalgebras.
 
It would be rewarding to generalize the concrete example to other supersymmetric integrable hierarchies and superalgebras. In our previous work on the bosonic theory a map between Lifshitz solution of $sl(N,\mathbb{R})$ Chern-Simons theory and Lifshitz scaling exponent $z$  to the $(m,n)$ member of the KdV hierarchy where $m=z$ and $n=N$ was established by Drinfeld-Sokolov \cite{Drinfeld:1984qv}
 formalism. It would be very interesting to see whether such a map with different Lifshitz scaling exponent can be constructed  for the supersymmetric theory using a supersymmetric generalization of Drinfeld-Sokolov formalism. If such a more general relation exists,  then it seems Lifshitz Chern-Simons theory provides a natural representation of the KdV hierarchy. Since supersymmetric 
 versions of the KdV and other integrable hierarchies are less studied the Chern-Simons  approach might be useful for a more systematic study. 
 
 Another possible direction for research lies in the the construction of black hole solutions in supersymmetric Chern-Simons Lifshitz theories following the work on the bosonic case \cite{Gutperle:2013oxa,Perez:2014pya}

%%%%%%%%%%%%%%%%%%%%%%%%%%%%%%%%%%%%%%%%%%%%
%%%%%%%%%%%%%%%%%%%%%%%%%%%%%%%%%%%%%%%%%%%%
\section*{ Acknowledgements}

The work of M. Gutperle and Yi Li is supported in part by the National Science Foundation under grant PHY-16-19926.  In addition Yi Li is  grateful to the Bhaumik Institute for theoretical Physics for support.

\newpage

\appendix

\section{A review of $sl(3|2)$  }
The bosonic part of $sl(3|2)$ is $U(1)\oplus sl(2,\mathbb{R})\oplus sl(3,\mathbb{R})$, it's generated by spin $2$ generators $L_i,A_i,\quad i=-1,0,1$, spin $3$ generators $W_i,\quad i=-2,-1,0,1,2$, and $J$ the generator of $u(1)$. The fermionic part of $sl(3|2)$ is generated by spin $\frac{3}{2}$ generators $G_r,H_r,\quad r=-\frac{1}{2},\frac{1}{2}$ and spin $\frac{5}{2}$ generators $S_r,T_r,\quad r=-\frac{3}{2},-\frac{1}{2},\frac{1}{2},\frac{3}{2}$. $L_i$ generate the $sl(2,\mathbb{R})$ subalgebra and the $L_0$ is the Cartan generator. The non-zero commutation relations are
\begin{align}
 &[L_i,L_j]=(i-j)L_{i+j} \quad [A_i,A_j]=(i-j)L_{i+j} \quad [L_i,A_j]=(i-j)A_{i+j} \nonumber \\
 &[L_i,W_j]=(2i-j)W_{i+j} \quad [A_i,W_j]=(2i-j)W_{i+j}  \nonumber  \\
 &[W_i,W_j]=\frac{1}{6}(j-i)(2i^2+2j^2-ij-8)(L_{i+j}+A_{i+j})  \nonumber \\
 &[L_i,G_r]=(\frac{i}{2}-r)G_{i+r} \quad [L_i,H_r]=(\frac{i}{2}-r)H_{i+r}  \nonumber  \\
 &[L_i,S_r]=(\frac{3i}{2}-r)S_{i+r} \quad [L_i,T_r]=(\frac{3i}{2}-r)T_{i+r}  \nonumber \\
 &[A_i,G_r]=\frac{4}{3}S_{i+r}+\frac{5}{3}(\frac{i}{2}-r)G_{i+r} \quad [A_i,H_r]=-\frac{4}{3}T_{i+r}+\frac{5}{3}(\frac{i}{2}-r)H_{i+r}  \nonumber \\
 &[A_i,S_r]=\frac{1}{3}(\frac{3i}{2}-r)S_{i+r}-\frac{1}{3}(3i^2-2ir+r^2-\frac{9}{4})G_{i+r} \nonumber  \\
 &[A_i,T_r]=\frac{1}{3}(\frac{3i}{2}-r)T_{i+r}+\frac{1}{3}(3i^2-2ir+r^2-\frac{9}{4})H_{i+r}  \nonumber \\
 &[W_i,G_r]=-\frac{4}{3}(\frac{i}{2}-2r)S_{i+r} \quad [W_i,H_r]=-\frac{4}{3}(\frac{i}{2}-2r)T_{i+r}  \nonumber \\
 &[W_i,S_r]=-\frac{1}{3}(2r^2-2ir+i^2-\frac{5}{2})S_{i+r}-\frac{1}{6}(4r^3-3ir^2+2i^2r-i^3-9r-\frac{19}{4}i)G_{i+r}  \nonumber \\
 &[W_i,T_r]=-\frac{1}{3}(2r^2-2ir+i^2-\frac{5}{2})T_{i+r}-\frac{1}{6}(4r^3-3ir^2+2i^2r-i^3-9r-\frac{19}{4}i)H_{i+r} \nonumber  \\
 &[J,G_r]=G_r \quad [J,H_r]=-H_r \quad [J,S_r]=S_r \quad[J,T_r]=-T_r  \nonumber \\
 &\{G_r,H_s\}=2L_{r+s}+(r-s)J \nonumber  \\
 &\{S_r,T_s\}=-\frac{3}{4}(r-s)W_{r+s}+\frac{1}{8}(3s^2-4rs+3r^2-\frac{9}{2})(L_{r+s}-3A_{r+s})-\frac{1}{4}(r-s)(r^2+s^2-\frac{5}{2})J  \nonumber \\
 &\{G_r,T_s\}=-\frac{3}{2}W_{r+s}+\frac{3}{4}(3r-s)A_{r+s}-\frac{5}{4}(3r-s)L_{r+s} \nonumber  \\
 &\{H_r,S_s\}=-\frac{3}{2}W_{r+s}-\frac{3}{4}(3r-s)A_{r+s}+\frac{5}{4}(3r-s)L_{r+s}
\end{align}
The subindex is the weight of the element, it's the eigenvalue of the commutator with $L_0$, and can be raised (lowered) by $L_1$ ($L_{-1}$). A weight-flipping automorphism exists
\begin{align}
 &J \rightarrow -J   \nonumber \\
 &L_0 \rightarrow -L_0, \quad L_1 \rightarrow L_{-1}, \quad L_{-1} \rightarrow L_1  \nonumber  \\
 &A_0 \rightarrow -A_0, \quad A_1 \rightarrow A_{-1}, \quad A_{-1} \rightarrow A_1 \nonumber  \\
 &W_2 \rightarrow -W_{-2}, \quad W_1 \rightarrow W_{-1}, \quad W_0 \rightarrow -W_0, \quad W_{-1} \rightarrow W_1, \quad W_{-2} \rightarrow -W_2  \nonumber  \\
 &G_\frac{1}{2} \rightarrow H_{-\frac{1}{2}}, \quad G_{-\frac{1}{2}} \rightarrow -H_\frac{1}{2}  \nonumber \\
 &H_\frac{1}{2} \rightarrow -G_{-\frac{1}{2}}, \quad H_{-\frac{1}{2}} \rightarrow G_\frac{1}{2}  \nonumber  \\
 &S_\frac{3}{2} \rightarrow -T_{-\frac{3}{2}}, \quad S_\frac{1}{2} \rightarrow T_{-\frac{1}{2}}, \quad S_{-\frac{1}{2}} \rightarrow  -T_\frac{3}{2}, \quad S_{-\frac{3}{2}} \rightarrow T_\frac{3}{2}  \nonumber \\
 &T_\frac{3}{2} \rightarrow S_{-\frac{3}{2}}, \quad T_\frac{1}{2} \rightarrow -S_{-\frac{1}{2}}, \quad T_{-\frac{1}{2}} \rightarrow S_\frac{1}{2}, \quad T_{-\frac{3}{2}} \rightarrow -S_\frac{3}{2}
\end{align}

The defining representation by super matrix is given by the following expressions

\begin{align}
 &J = \left( \begin{array}{ccccc}
 2 & 0 & 0 & 0 & 0 \\
 0 & 2 & 0 & 0 & 0 \\
 0 & 0 & 2 & 0 & 0 \\
 0 & 0 & 0 & 3 & 0 \\
 0 & 0 & 0 & 0 & 3 \\
 \end{array} \right) \nonumber  \\
 &L_0 = \left( \begin{array}{ccccc}
 1 & 0 & 0 & 0 & 0 \\
 0 & 0 & 0 & 0 & 0 \\
 0 & 0 & -1 & 0 & 0 \\
 0 & 0 & 0 & \frac{1}{2} & 0 \\
 0 & 0 & 0 & 0 & -\frac{1}{2} \\
 \end{array} \right),
 L_1 = \left( \begin{array}{ccccc}
 0 & 0 & 0 & 0 & 0 \\
 \sqrt{2} & 0 & 0 & 0 & 0 \\
 0 & \sqrt{2} & 0 & 0 & 0 \\
 0 & 0 & 0 & 0 & 0 \\
 0 & 0 & 0 & 1 & 0 \\
 \end{array} \right),
 L_{-1} = \left( \begin{array}{ccccc}
 0 & -\sqrt{2} & 0 & 0 & 0 \\
 0 & 0 & -\sqrt{2} & 0 & 0 \\
 0 & 0 & 0 & 0 & 0 \\
 0 & 0 & 0 & 0 & -1 \\
 0 & 0 & 0 & 0 & 0 \\
 \end{array} \right)\nonumber  \\
 &A_0 = \left( \begin{array}{ccccc}
 1 & 0 & 0 & 0 & 0 \\
 0 & 0 & 0 & 0 & 0 \\
 0 & 0 & -1 & 0 & 0 \\
 0 & 0 & 0 & -\frac{1}{2} & 0 \\
 0 & 0 & 0 & 0 & \frac{1}{2} \\
 \end{array} \right),
 A_1 = \left( \begin{array}{ccccc}
 0 & 0 & 0 & 0 & 0 \\
 \sqrt{2} & 0 & 0 & 0 & 0 \\
 0 & \sqrt{2} & 0 & 0 & 0 \\
 0 & 0 & 0 & 0 & 0 \\
 0 & 0 & 0 & -1 & 0 \\
 \end{array} \right),
 A_{-1} = \left( \begin{array}{ccccc}
 0 & -\sqrt{2} & 0 & 0 & 0 \\
 0 & 0 & -\sqrt{2} & 0 & 0 \\
 0 & 0 & 0 & 0 & 0 \\
 0 & 0 & 0 & 0 & 1 \\
 0 & 0 & 0 & 0 & 0 \\
 \end{array} \right)  \nonumber  \\
 &W_2 = \left( \begin{array}{ccccc}
 0 & 0 & 0 & 0 & 0 \\
 0 & 0 & 0 & 0 & 0 \\
 4 & 0 & 0 & 0 & 0 \\
 0 & 0 & 0 & 0 & 0 \\
 0 & 0 & 0 & 0 & 0 \\
 \end{array} \right),
 W_{-2} = \left( \begin{array}{ccccc}
 0 & 0 & 4 & 0 & 0 \\
 0 & 0 & 0 & 0 & 0 \\
 0 & 0 & 0 & 0 & 0 \\
 0 & 0 & 0 & 0 & 0 \\
 0 & 0 & 0 & 0 & 0 \\
 \end{array} \right)\nonumber
\end{align}
\begin{align}
 &W_1 = \left( \begin{array}{ccccc}
 0 & 0 & 0 & 0 & 0 \\
 \sqrt{2} & 0 & 0 & 0 & 0 \\
 0 & -\sqrt{2} & 0 & 0 & 0 \\
 0 & 0 & 0 & 0 & 0 \\
 0 & 0 & 0 & 0 & 0 \\
 \end{array} \right),
 W_{-1} = \left( \begin{array}{ccccc}
 0 & -\sqrt{2} & 0 & 0 & 0 \\
 0 & 0 & \sqrt{2} & 0 & 0 \\
 0 & 0 & 0 & 0 & 0 \\
 0 & 0 & 0 & 0 & 0 \\
 0 & 0 & 0 & 0 & 0 \\
 \end{array} \right),
 W_0= \left( \begin{array}{ccccc}
 \frac{2}{3} & 0 & 0 & 0 & 0 \\
 0 & -\frac{4}{3} & 0 & 0 & 0 \\
 0 & 0 & \frac{2}{3} & 0 & 0 \\
 0 & 0 & 0 & 0 & 0 \\
 0 & 0 & 0 & 0 & 0 \\
 \end{array} \right)\nonumber  \\
 &G_\frac{1}{2} = \left( \begin{array}{ccccc}
 0 & 0 & 0 & 0 & 0 \\
 0 & 0 & 0 & 0 & 0 \\
 0 & 0 & 0 & 0 & 0 \\
 2 & 0 & 0 & 0 & 0 \\
 0 & \sqrt{2} & 0 & 0 & 0 \\
 \end{array} \right),
 G_{-\frac{1}{2}} = \left( \begin{array}{ccccc}
 0 & 0 & 0 & 0 & 0 \\
 0 & 0 & 0 & 0 & 0 \\
 0 & 0 & 0 & 0 & 0 \\
 0 & -\sqrt{2} & 0 & 0 & 0 \\
 0 & 0 & -2 & 0 & 0 \\
 \end{array} \right)\nonumber \\
 &H_\frac{1}{2} = \left( \begin{array}{ccccc}
 0 & 0 & 0 & 0 & 0 \\
 0 & 0 & 0 & \sqrt{2} & 0 \\
 0 & 0 & 0 & 0 & 2 \\
 0 & 0 & 0 & 0 & 0 \\
 0 & 0 & 0 & 0 & 0 \\
 \end{array} \right),
 H_{-\frac{1}{2}} = \left( \begin{array}{ccccc}
 0 & 0 & 0 & 2 & 0 \\
 0 & 0 & 0 & 0 & \sqrt{2} \\
 0 & 0 & 0 & 0 & 0 \\
 0 & 0 & 0 & 0 & 0 \\
 0 & 0 & 0 & 0 & 0 \\
 \end{array} \right)\nonumber \\
 &S_\frac{3}{2} = \left( \begin{array}{ccccc}
 0 & 0 & 0 & 0 & 0 \\
 0 & 0 & 0 & 0 & 0 \\
 0 & 0 & 0 & 0 & 0 \\
 0 & 0 & 0 & 0 & 0 \\
 -3 & 0 & 0 & 0 & 0 \\
 \end{array} \right),
 S_{-\frac{3}{2}} = \left( \begin{array}{ccccc}
 0 & 0 & 0 & 0 & 0 \\
 0 & 0 & 0 & 0 & 0 \\
 0 & 0 & 0 & 0 & 0 \\
 0 & 0 & -3 & 0 & 0 \\
 0 & 0 & 0 & 0 & 0 \\
 \end{array} \right)\nonumber \\
 &S_\frac{1}{2} = \left( \begin{array}{ccccc}
 0 & 0 & 0 & 0 & 0 \\
 0 & 0 & 0 & 0 & 0 \\
 0 & 0 & 0 & 0 & 0 \\
 -1 & 0 & 0 & 0 & 0 \\
 0 & \sqrt{2} & 0 & 0 & 0 \\
 \end{array} \right),
 S_{-\frac{1}{2}} = \left( \begin{array}{ccccc}
 0 & 0 & 0 & 0 & 0 \\
 0 & 0 & 0 & 0 & 0 \\
 0 & 0 & 0 & 0 & 0 \\
 0 & \sqrt{2} & 0 & 0 & 0 \\
 0 & 0 & -1 & 0 & 0 \\
 \end{array} \right)\nonumber \\
 &T_\frac{3}{2} = \left( \begin{array}{ccccc}
 0 & 0 & 0 & 0 & 0 \\
 0 & 0 & 0 & 0 & 0 \\
 0 & 0 & 0 & -3 & 0 \\
 0 & 0 & 0 & 0 & 0 \\
 0 & 0 & 0 & 0 & 0 \\
 \end{array} \right),
 T_{-\frac{3}{2}} = \left( \begin{array}{ccccc}
 0 & 0 & 0 & 0 & 3 \\
 0 & 0 & 0 & 0 & 0 \\
 0 & 0 & 0 & 0 & 0 \\
 0 & 0 & 0 & 0 & 0 \\
 0 & 0 & 0 & 0 & 0 \\
 \end{array} \right)\nonumber \\
 &T_\frac{1}{2} = \left( \begin{array}{ccccc}
 0 & 0 & 0 & 0 & 0 \\
 0 & 0 & 0 & -\sqrt{2} & 0 \\
 0 & 0 & 0 & 0 & 1 \\
 0 & 0 & 0 & 0 & 0 \\
 0 & 0 & 0 & 0 & 0 \\
 \end{array} \right),
 T_{-\frac{1}{2}} = \left( \begin{array}{ccccc}
 0 & 0 & 0 & -1 & 0 \\
 0 & 0 & 0 & 0 & \sqrt{2} \\
 0 & 0 & 0 & 0 & 0 \\
 0 & 0 & 0 & 0 & 0 \\
 0 & 0 & 0 & 0 & 0 \\
 \end{array} \right)
\end{align}
They are all super-traceless, and closed under multiplication with the identity super matrix added. In addition, the weight is additive under super matrix multiplication if we count the weight of the identity super matrix as zero. In this super matrix representation, the weight-flipping automorphism is simply given by taking the negative of the transposition of the bosonic elements and the transposition of the fermionic elements.

\section{Poisson structure of $sl(3|2)$ Chern-Simons theory}

As an example, we show how to calculate the Poisson bracket $\{h(x^\prime),g(x)\}$. Clearly we need to find a gauge transformation parameter $\lambda$ which is associated with the charge $Q_\lambda$ that takes the form of an integral of the product of $h$ and an arbitrary fermionic function. $\rm str(G_\frac{1}{2}H_{-\frac{1}{2}})$ is nonzero so we want $\lambda$ to start with $\gamma G_\frac{1}{2}$, where $\gamma$ is an arbitrary fermionic function. The other non-highest weight terms in $\lambda$ are determined by requiring the gauge transformation preserves the lowest weight gauge of $a_x$ and we find
\begin{align}
 \lambda = \gamma G_\frac{1}{2} - (\gamma^{'}+\gamma j)G_{-\frac{1}{2}} - \frac{9}{4} \gamma a S_{-\frac{3}{2}} + \gamma h L_{-1} - \frac{3}{8} \gamma t W_{-2}
\end{align}
The associated charge is calculated as
\begin{align}
 \delta Q_\lambda & = \frac{1}{6\alpha} \int dx\; {\rm str}\; (\lambda \delta a_x) = -\frac{1}{\alpha} \int dx \gamma \delta  \nonumber  \\
 Q_\lambda &= -\frac{1}{\alpha} \int dx \gamma h
\end{align}
The gauge transformation on $g$ is calculated to be
\begin{align}
 &\delta_\lambda g(x) = -\gamma(x)(\frac{5}{3}a(x)+j(x)^2+l(x)+j^{'}(x))-2\gamma^{'}(x)j(x)-\gamma^{''}(x) \nonumber  \\
 &= \{Q_\lambda,g(x)\} = -\frac{1}{\alpha}\int dx^\prime \gamma(x^\prime) \{h(x^\prime),g(x)\}
\end{align}
Therefore
\begin{align}
 \{h(x^\prime),g(x)\}= \alpha(\frac{5}{3}a(x)+j(x)^2+l(x)+j^{'}(x))\delta(x^\prime-x) - 2\alpha j(x)\delta^{'}(x^\prime-x) + \alpha \delta^{''}(x^\prime-x)
\end{align}

We list the Poisson brackets of all the fields here
\begin{align}
 &\{j,j\}=\alpha \delta^{'} \nonumber  \\
 &\{j,g\}=\alpha g \delta  \nonumber \\
 &\{j,h\}=-\alpha h \delta \nonumber  \\
 &\{j,s\}=\alpha s \delta  \nonumber \\
 &\{j,t\}=-\alpha t \delta \nonumber  \\
 &\{h,g\}=\alpha(\frac{5}{3}a+j^2+l+j^{'})\delta - 2\alpha j \delta^{'} + \alpha \delta^{''} \nonumber  \\
 &\{h,s\}=\frac{4}{9}\alpha(-6w+a^{'}+4aj)\delta - \frac{16}{9}\alpha a \delta^{'}  \nonumber \\
 &\{h,a\}=\frac{9}{4}\alpha t \delta \nonumber  \\
 &\{h,l\}=-\alpha(2jh+h^{'}+\frac{15}{4}t)\delta+3\alpha h \delta^{'} \nonumber  \\
 &\{h,w\}=\alpha(\frac{2}{3}ah+\frac{3}{2}jt+\frac{3}{8}t^{'})\delta + \frac{15}{8}\alpha t \delta^{'}  \nonumber \\
 &\{g,t\}=-\frac{4}{9}\alpha(6w+a^{'}-4aj)\delta + \frac{16}{9}\alpha a \delta^{'}  \nonumber \\
 &\{g,a\}=-\frac{9}{4}\alpha s \delta  \nonumber  \\
 &\{g,l\}=-\alpha(-2jg+g^{'}-\frac{15}{4}t)\delta+3\alpha g \delta^{'}  \nonumber \\
 &\{g,w\}=-\alpha(\frac{2}{3}ag+\frac{3}{2}js-\frac{3}{8}s^{'})\delta - \frac{15}{8}\alpha s \delta^{'}  \nonumber \\
 &\frac{1}{6\alpha}\{5a+3l+3j^2,g\}=-g^{'}\delta + \frac{3}{2}g\delta^{'}  \nonumber \\
 &\frac{1}{6\alpha}\{5a+3l+3j^2,h\}=-h^{'}\delta + \frac{3}{2}h\delta^{'}  \nonumber \\
 &\frac{1}{6\alpha}\{5a+3l+3j^2,a\}=-a^{'}\delta + 2a\delta^{'}  \nonumber \\
 &\frac{1}{6\alpha}\{5a+3l+3j^2,l\}=-l^{'}\delta + 2l\delta^{'} + \frac{1}{2}\delta^{'''}  \nonumber \\
 &\frac{1}{6\alpha}\{5a+3l+3j^2,s\}=-s^{'}\delta + \frac{5}{2}s\delta^{'} \nonumber  \\
 &\frac{1}{6\alpha}\{5a+3l+3j^2,t\}=-t^{'}\delta + \frac{5}{2}t\delta^{'}  \nonumber \\
 &\frac{1}{6\alpha}\{5a+3l+3j^2,w\}=-w^{'}\delta + 3w\delta^{'} \nonumber 
\end{align}
\begin{align}
 &\{a,a\}=-\frac{3}{8}\alpha(5a^{'}-3l^{'})\delta + \frac{3}{8}\alpha(10a-6l)\delta^{'} - \frac{9}{16}\alpha \delta^{'''} \nonumber  \\
 &\{a,s\}=-\frac{\alpha}{4}(g(-9a+9l+j^2+j^{'})+6js+2jg^{'}+6s^{'}+g^{''})\delta + \frac{\alpha}{4}(4jg+15s+4g^{'})\delta^{'} - \frac{3\alpha}{2}g\delta^{''} \nonumber  \\
 &\{a,t\}=-\frac{\alpha}{4}(h(9a-9l-j^2+j^{'})-6jt+2jh^{'}+6t^{'}-h^{''})\delta + \frac{\alpha}{4}(4jh+15t-4h^{'})\delta^{'} + \frac{3\alpha}{2}h\delta^{''} \nonumber  \\
 &\{a,w\}=-\frac{3\alpha}{16}(-15(gt+hs)+5(gh)^{'}+4w^{'})\delta + \frac{3\alpha}{16}(15gh+12w)\delta^{'} \nonumber  \\
 &\{t,t\}=\frac{2\alpha}{3}(10ht+\frac{4}{3}hh^{'})\delta  \nonumber \\
 &\{s,s\}=-\frac{2\alpha}{3}(10gs-\frac{4}{3}gg^{'})\delta  \nonumber \\
 &\{t,s\}=-\frac{2\alpha}{3}(-\frac{5}{2}a^2+\frac{5}{9}aj^2+\frac{1}{6}j^4+al+\frac{5}{3}j^2l+\frac{3}{2}l^2+\frac{8}{3}jgh+4gt+3gh^{'}-4hs-\frac{4}{3}hg^{'}-\frac{8}{3}jw+\frac{5}{9}(ja)^{'}  \nonumber \\
 &+\frac{5}{3}(jl)^{'}+j^2j^{'}+\frac{1}{2}(j^{'})^2-\frac{4}{3}w^{'}+\frac{1}{6}a^{''}+\frac{2}{3}jj^{''}+\frac{1}{2}l^{''}+\frac{1}{6}j^{'''})\delta  \nonumber  \\
 &+ \frac{2\alpha}{3}(\frac{10}{9}aj+\frac{2}{3}j^3+\frac{10}{3}jl+\frac{13}{3}gh)-\frac{8}{3}w+\frac{5}{9}a+2jj^{'}+\frac{5}{3}l^{'}+\frac{2}{3}j^{''})\delta^{'}  \nonumber \\
 &-\frac{2\alpha}{3}(\frac{5}{9}a+j^2+\frac{5}{3}l+j^{'})\delta^{''}+\frac{4\alpha}{9}\delta^{'''}-\frac{\alpha}{9}\delta^{(4)}  \nonumber  \\
 &\{t,w\}=-\frac{2\alpha}{3}(\frac{13}{12}ajh+\frac{1}{4}j^3h+\frac{9}{4}jlh-\frac{33}{8}at-\frac{3}{8}j^2t-\frac{15}{8}lt-5wh+\frac{13}{16}a^{'}h+\frac{13}{16}ah^{'}+\frac{3}{16}j^2h^{'}+\frac{19}{16}lh^{'}  \nonumber  \\
 &+\frac{9}{8}jj^{'}h-\frac{3}{4}j^{'}t+\frac{7}{16}j^{'}h^{'}+\frac{27}{16}l^{'}h-\frac{3}{8}jt^{'}+\frac{1}{8}jh^{''}+\frac{9}{16}j^{''}h-\frac{3}{16}t^{''}+\frac{1}{16}h^{'''})\delta \nonumber  \\
 &+ \frac{2\alpha}{3}(\frac{91}{48}ah+\frac{15}{16}j^2h+\frac{55}{16}lh-\frac{9}{8}jt+\frac{5}{8}jh^{'}+\frac{25}{16}j^{'}h-\frac{3}{4}t^{'}+\frac{5}{16}h^{''})\delta^{'}  \nonumber \\
 &- \frac{2\alpha}{3}(\frac{5}{4}jh-\frac{15}{16}t+\frac{5}{8}h^{'})\delta^{''} + \frac{5}{12}\alpha h \delta^{'''}  \nonumber \\
 &\{s,w\}=-\frac{2\alpha}{3}(-\frac{13}{12}ajg-\frac{1}{4}j^3g-\frac{9}{4}jlg+\frac{33}{8}as+\frac{3}{8}j^2s+\frac{15}{8}ls+5wg+\frac{13}{16}a^{'}g+\frac{13}{16}ag^{'}+\frac{3}{16}j^2g^{'}+\frac{19}{16}lg^{'}  \nonumber  \\
 &+\frac{9}{8}jj^{'}g-\frac{3}{4}j^{'}s-\frac{7}{16}j^{'}g^{'}+\frac{27}{16}l^{'}g-\frac{3}{8}js^{'}-\frac{1}{8}jg^{''}-\frac{9}{16}j^{''}g+\frac{3}{16}s^{''}+\frac{1}{16}g^{'''})\delta  \nonumber  \\
 &+ \frac{2\alpha}{3}(\frac{91}{48}ag+\frac{15}{16}j^2g+\frac{55}{16}lg-\frac{9}{8}js-\frac{5}{8}jg^{'}-\frac{25}{16}j^{'}g+\frac{3}{4}s^{'}+\frac{5}{16}g^{''})\delta^{'}  \nonumber \\
 &- \frac{2\alpha}{3}(-\frac{5}{4}jg+\frac{15}{16}s+\frac{5}{8}g^{'})\delta^{''} + \frac{5}{12}\alpha g \delta^{'''}  \nonumber  \\
 &\{w,w\}=\frac{\alpha}{32}(33(gt)^{'}-33(hs)^{'}-28(jgh)^{'}+14(gh^{''}-g^{''}h)+16((a+l)^2)^{'}+2(a+l)^{'''})\delta \nonumber  \\
 &-\frac{\alpha}{32}(32(a+l)^2-56jgh+66(gt-hs)+28gh^{'}-28g^{'}h+9(a+l)^{''})\delta^{'}  \nonumber \\
 &+\frac{15\alpha}{32}(a+l)^{'}\delta^{''} - \frac{5\alpha}{16}(a+l)\delta^{'''} - \frac{\alpha}{64} \delta^{(5)}
\end{align}
where the first field in the bracket is at $x^\prime$ and the second is at $x$, all the fields on the right hand side are at $x$, and $\delta$ is short for $\delta(x^\prime-x)$. Brackets of fields not listed above are either zero or can be inferred from the brackets listed by simple principle, for example, antisymmetry of Poisson brackets.

\newpage
 %%%%%%%%%%%%%%%%%%%%%%%%%
 %%% Bibliography
 %%%%%%%%%%%%%%%%%%%%%%%%%

\end{document}